\input{aipcheck}

\documentclass[
    ,final            
  ]
  {aipproc}

\layoutstyle{6x9}

\usepackage{rotating}

\begin{document}

\def\met{{E^{\mathrm{miss}}_\mathrm{T}}}

\title{Determination of QCD Backgrounds in ATLAS: A challenge for SUSY searches}


\author{Bernhard Meirose \\ 
On behalf of the ATLAS Collaboration
}{
  address={Albert-Ludwigs-Universit\"{a}t Freiburg \\
Hermann-Herder-Str. 3 79104 Freiburg i. Br., Germany}
}

\begin{abstract}
In this paper we briefly discuss the estimation of uncertainties in QCD backgrounds to searches for Supersymmetry under development by the ATLAS collaboration. 
\end{abstract}

\maketitle


\section{Introduction}

The understanding of QCD jet events is the dominant background-determination challenge in Supersymmetry (SUSY) searches, in events containing jets and missing transverse energy ($\met$). Effects such as jet punch through the calorimeters and cosmic ray backgrounds can affect the $\met$. Other effects like pile-up of cavern and beam halo backgrounds can dramatically increase the cross-section of high $\met$ QCD backgrounds. Sources of systematic uncertainties in the estimation of QCD backgrounds include sensitivity to underlying event models and parton distribution functions. In the next sections we discuss important techniques for estimating the QCD backgrounds uncertainties in searches for SUSY, under development by the ATLAS \cite{ATLAS} collaboration. We evaluate the effects in terms of the effective mass, here defined as $\mathrm{M_{eff}} = \sum_{i=1}^{4} |p_T(j_i)| + \met $, for its important discriminative power for SUSY. All studies shown here are for 14 TeV center-of-mass energy, but the methods are also valid at lower energies.

 

\section{Jet Energy Scale}

QCD background estimates from Monte Carlo, contribute a large irreducible uncertainty to $\mathrm{M_{eff}}$ measurement, which derives from the uncertainty in the jet-energy-scale (JES). These effects
are estimated in what follows. For each event, the energy and momentum of each jet has been scaled by a constant factor, corresponding to a 10\%, 5\%, 3\% and 1\% uncertainty on the JES. The $\met$ has also been adjusted accordingly, by an amount corresponding to the change in the sum of the jet momenta in the x and y directions. Figure 1 shows the resulting $\mathrm{M_{eff}}$ distribution. The shaded bands show the uncertainties on the distribution for the four different values of the JES uncertainty considered. For a 10\% value, the uncertainty on the $\mathrm{M_{eff}}$ distribution ranges from $\sim$ 50-150\%. An improved understanding of the JES to a level of 5\% reduces the uncertainty on  $\mathrm{M_{eff}}$ by more than a factor of two, while a 3\% value shows an improved uncertainty on $\mathrm{M_{eff}}$ of between $\sim$ 10-30\%. If 1\% uncertainty in JES is achievable, the uncertainty in $\mathrm{M_{eff}}$ will be $\sim$ 5-10\%.

\begin{figure}
\begin{tabular}{c c}
  \includegraphics[height=.23\textheight,angle=0]{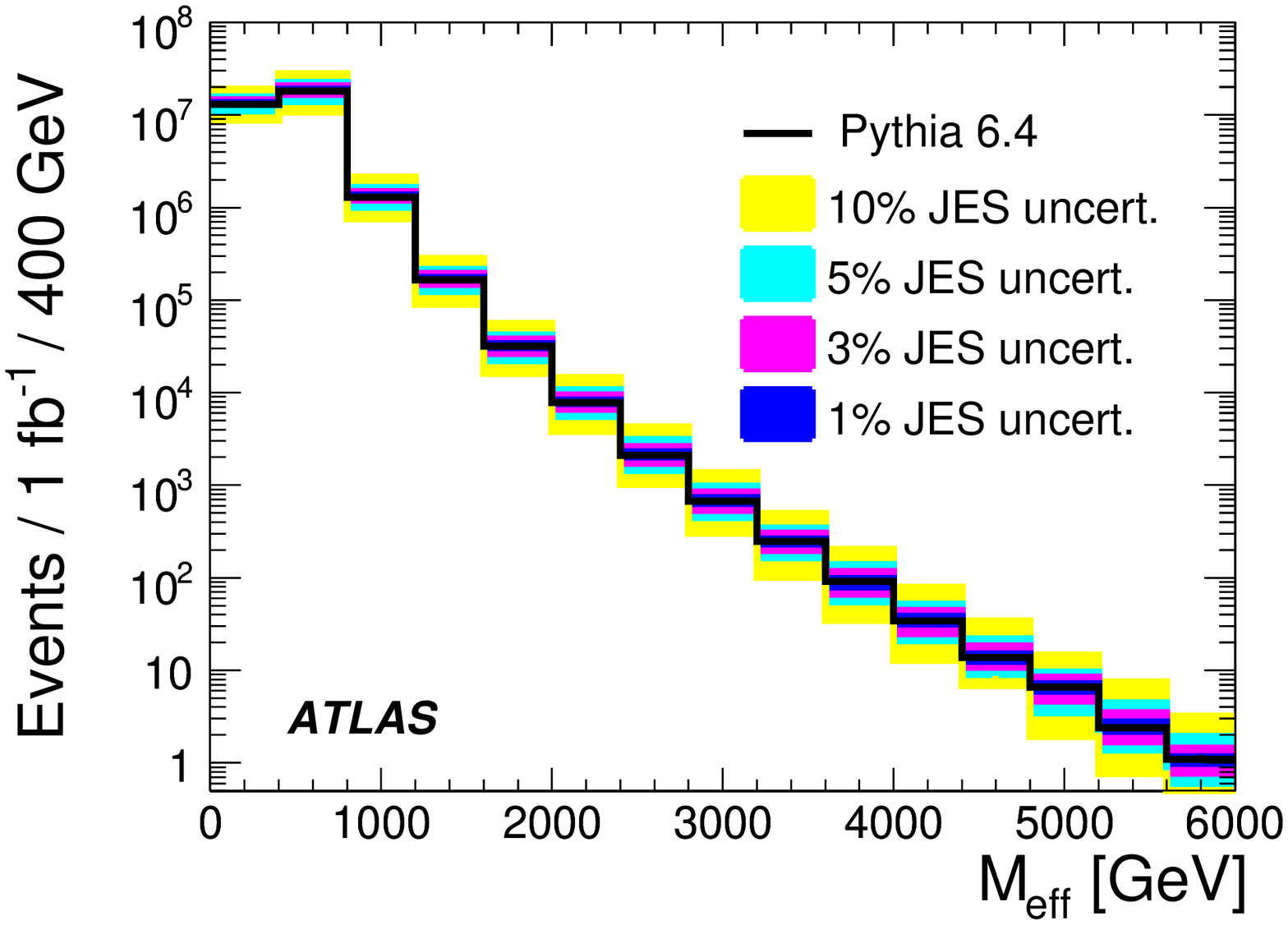}
&

  \includegraphics[height=.23\textheight,angle=0]{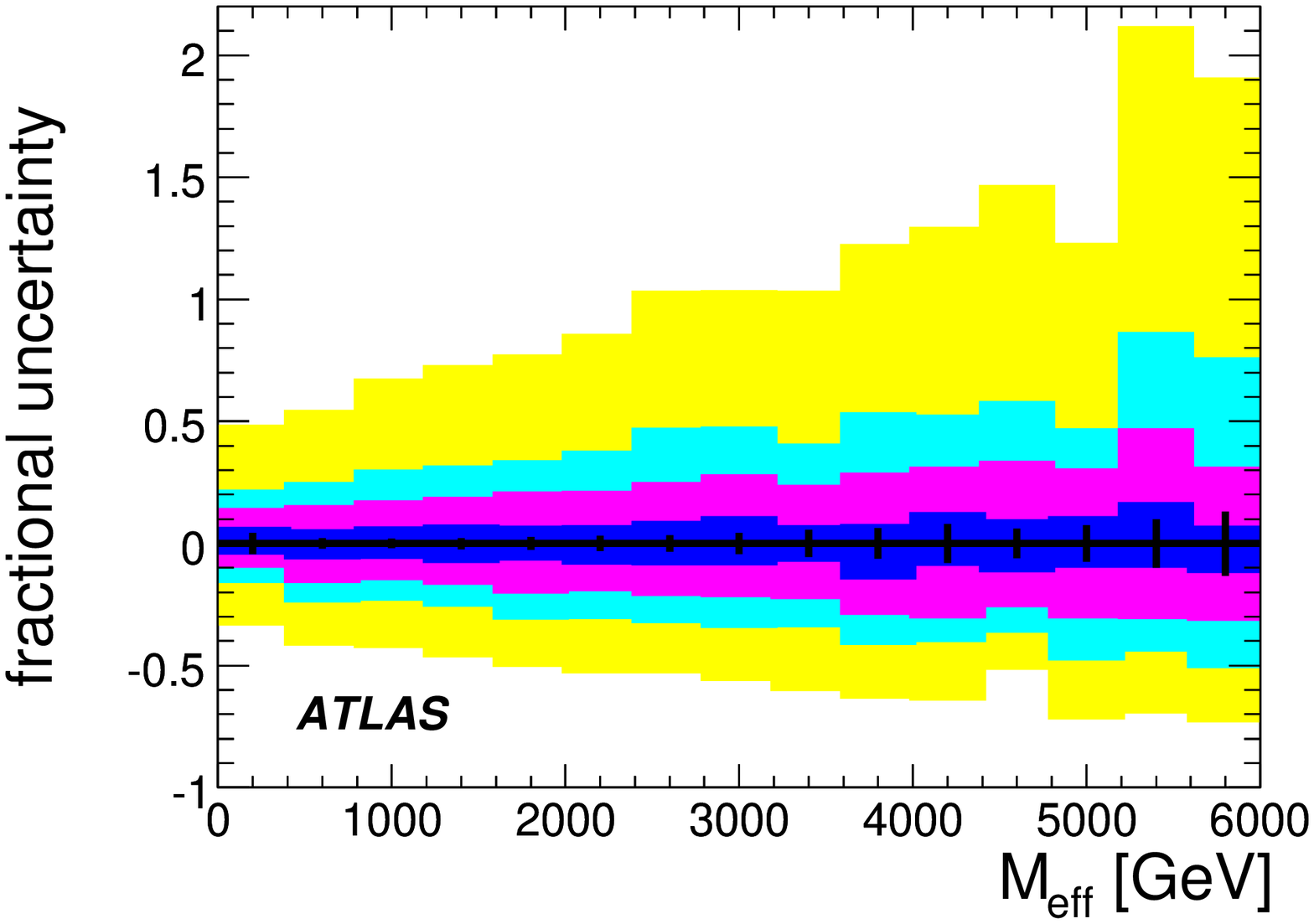}
\end{tabular}
  \caption{The $\mathrm{M_{eff}}$ distribution for simulated QCD events satisfying the ATLAS SUSY jet cuts \cite{CSC}. The left plot shows the number of events per bin and the right plot shows the fractional uncertainty with
respect to the central prediction. The shaded bands show the estimated uncertainty on the observable for
assumed JES uncertainties of 10\% (light), 5\% (medium-light), 3\% (medium) and 1\% (dark).
}
\end{figure}

\section{Generator Comparison}

In this section we compare systematic uncertainties arising from Monte Carlo generators. For comparison we use 
the leading-order ALPGEN 2.05 \cite{ALPGEN} code and the conventional stand-alone parton-shower generator PYTHIA 6.403 \cite{PYTHIA}. 

PYTHIA events were generated in different transverse-momentum ($p_T$) ranges of the two leading partons to study the high energy tails with sufficient statistics. Such sliced-sample production is also possible in ALPGEN. The number of PYTHIA events passing the ATLAS SUSY jet selection cuts \cite{CSC} is 2.1 times larger than the number of ALPGEN events for the same integrated luminosity. Both samples were normalized to 1 fb$^{-1}$ and ALPGEN samples were further multiplied by a factor of 2.1. Error bars on each histogram are based on the Monte Carlo statistics used in this study.

Figure 2 (left) shows the $p_T$ distribution of the leading jet for ALPGEN and PYTHIA events. In PYTHIA, for example the 2 $\rightarrow$ 2 scattering processes give softer leading $p_T$ jets because of the emmission of additional partons from the leading ones. The $\mathrm{M_{eff}}$ distribution between generators are shown in Figure 2, right. The $\mathrm{M_{eff}}$ distribution of ALPGEN events is harder than that of PYTHIA events, as expected from Figure 2, left.

\begin{figure}
\begin{tabular}{c c}
  \includegraphics[height=.23\textheight,angle=0]{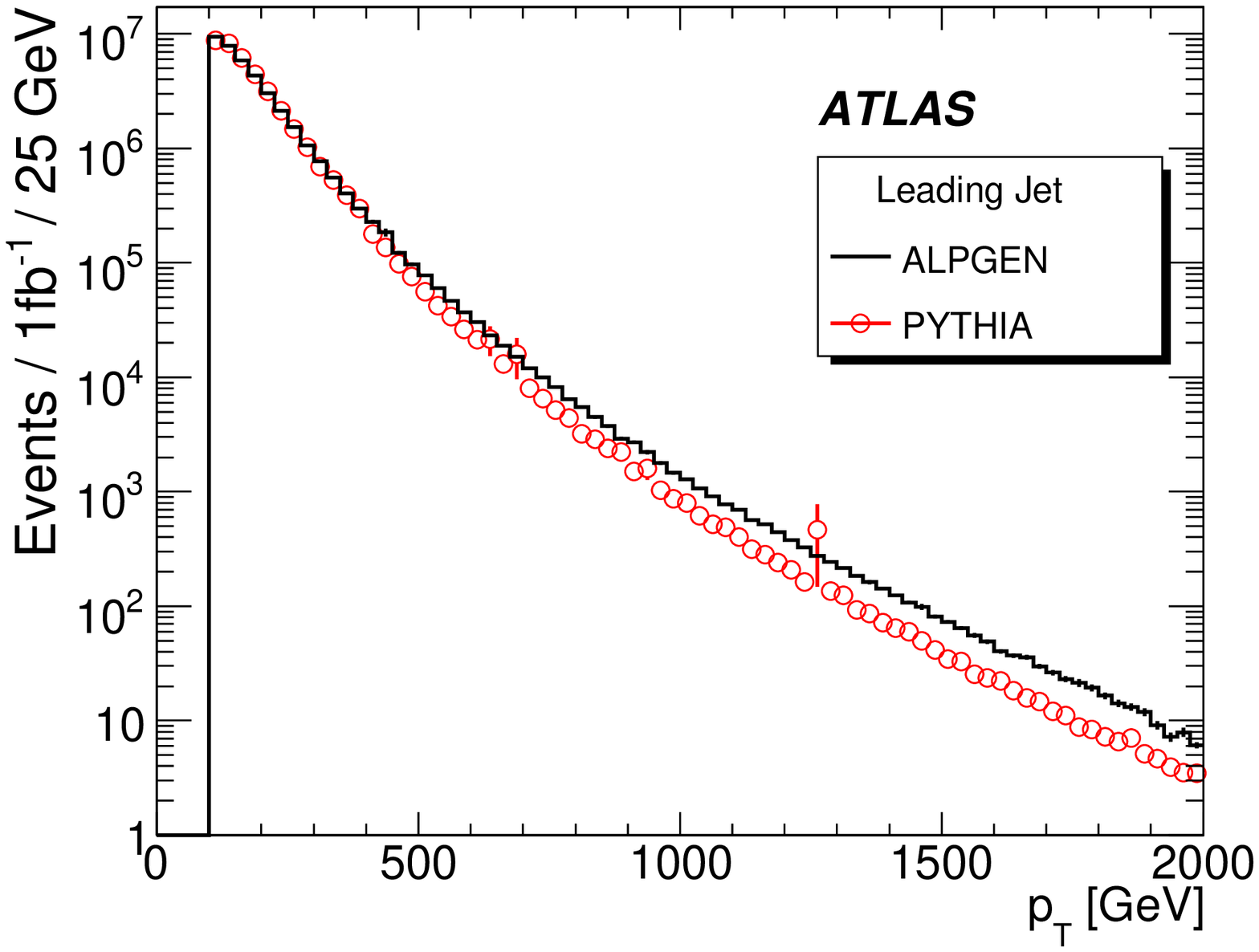}
&

  \includegraphics[height=.23\textheight,angle=0]{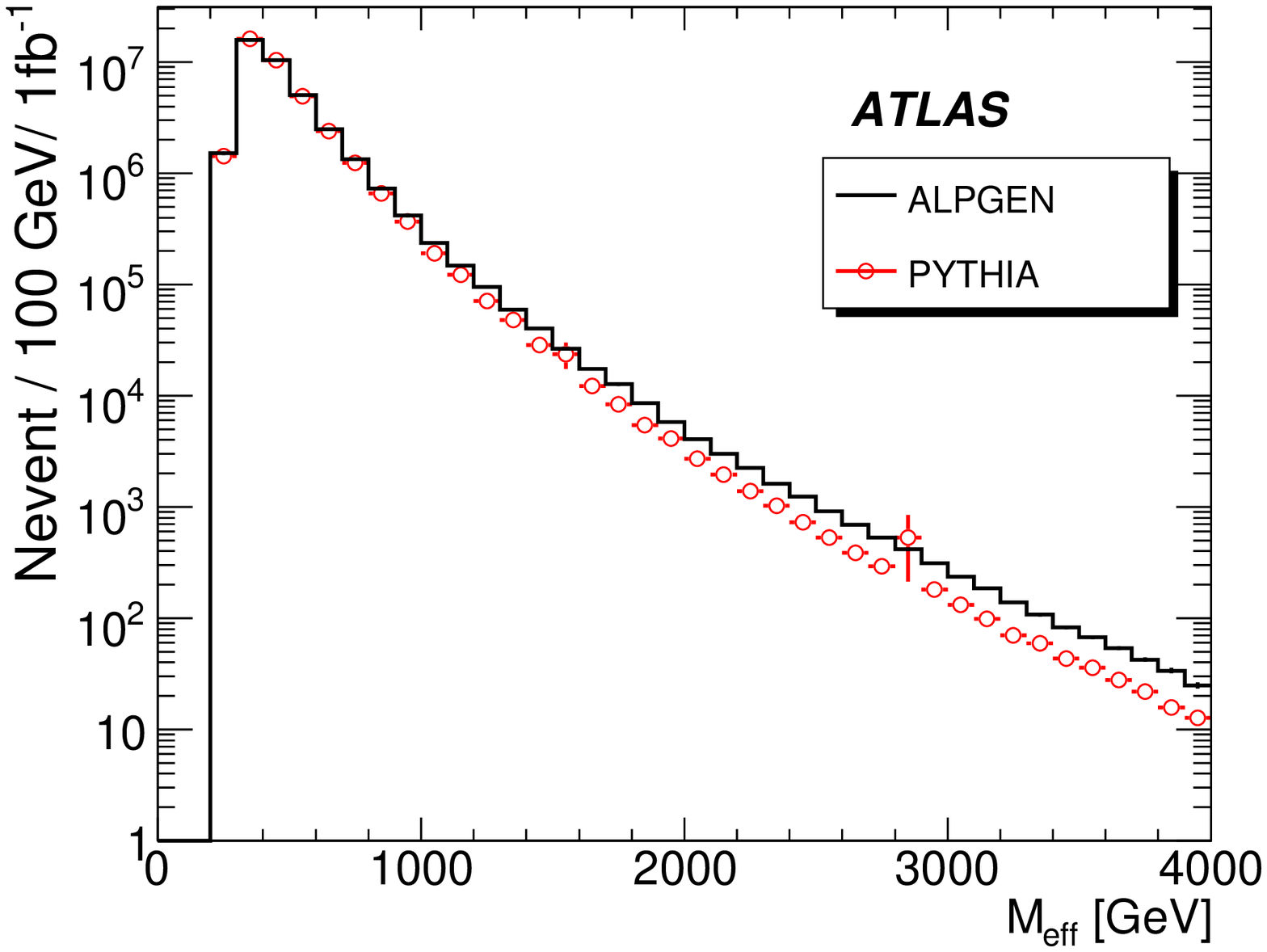}
\end{tabular}
  \caption{The $p_T$ of the leading jet (left) and ATLFAST \cite{ATLFAST} $\mathrm{M_{eff}}$ (right) distributions for PYTHIA (open circles) and ALPGEN (histogram) events.
}

\end{figure}

\section{Data Driven Background Estimates}

The inherent systematic and statistical uncertainties on Monte Carlo based QCD background estimates, limit their use in early LHC \cite{LHC} running. The understanding of the ATLAS detector and the underlying physics of QCD processes at high energy, will require a sufficient amount of data. For this reason data-driven background estimates with minimal reliance on Monte Carlo simulation will be a priority for the early phase of data-taking. The data driven approach in ATLAS relies on smearing the jet transverse momenta in regions of low $\met$ using a data-measured response function. This response function is defined as the distribution of event-by-event ratios of measured jet $p_T$ to true jet $p_T$ . The method to measure the calorimeter response to jets from data is divided in three steps: Gaussian response function measurement, full response function measurement including non-Gaussian tails and finally jet $p_T$ smearing.

The measurement of the gaussian response function relies on the "$\met$ projection method". In this procedure the gaussian response is extracted from the transverse momentum conservation in the $\gamma$ + jet events (photon-jet $p_T$ balance is required). 

For the non-gaussian response, events with the  $\met$ vector unambiguously associated to the jet in $\phi$, are used. Finally one combines the Gaussian and non-Gaussian components of the jet response measured. The full jet response function is plotted in Fig. 3, left.

The final step of the method consists of generating the events in a QCD Monte Carlo simulation using the full Gaussian + non-Gaussian response function to smear the jet transverse momenta in multijet events with low $\met$. These are referred to below as 'seed' events. Smeared events are constructed from each selected seed event by smearing the transverse momenta of their constituent jets with the full jet response function determined in the second step. The $\met$ of smeared events is then calculated by replacing, in the $\met$, the contribution from the $p_T$ of the 'seed' jets, by the one from the $p_T$ of the equivalent smeared jets.

Fig. 3 on the right shows the $\mathrm{M_{eff}}$ distribution for 23.8 pb$^{-1}$ of GEANT4 \cite{GEANT4} simulated ``data'' compared to smeared 'seed' events (QCD estimated) passing the ATLAS SUSY jet cuts \cite{CSC}. Good agreement can be seen between the estimated and GEANT4 ``data'' $\mathrm{M_{eff}}$ distributions. 

\begin{figure}
\begin{tabular}{c c}
  \includegraphics[height=.23\textheight,angle=0]{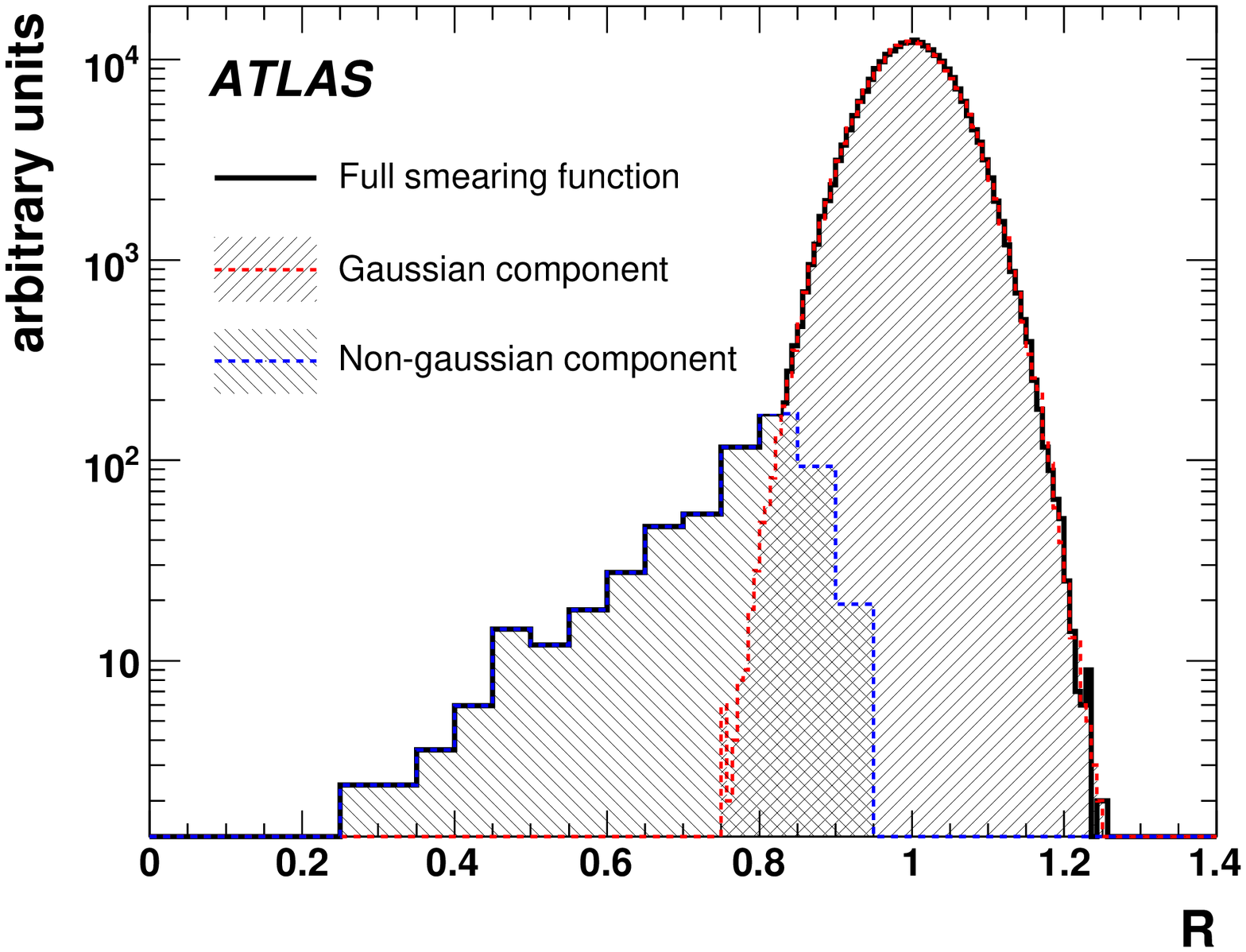}
&

  \includegraphics[height=.23\textheight,angle=0]{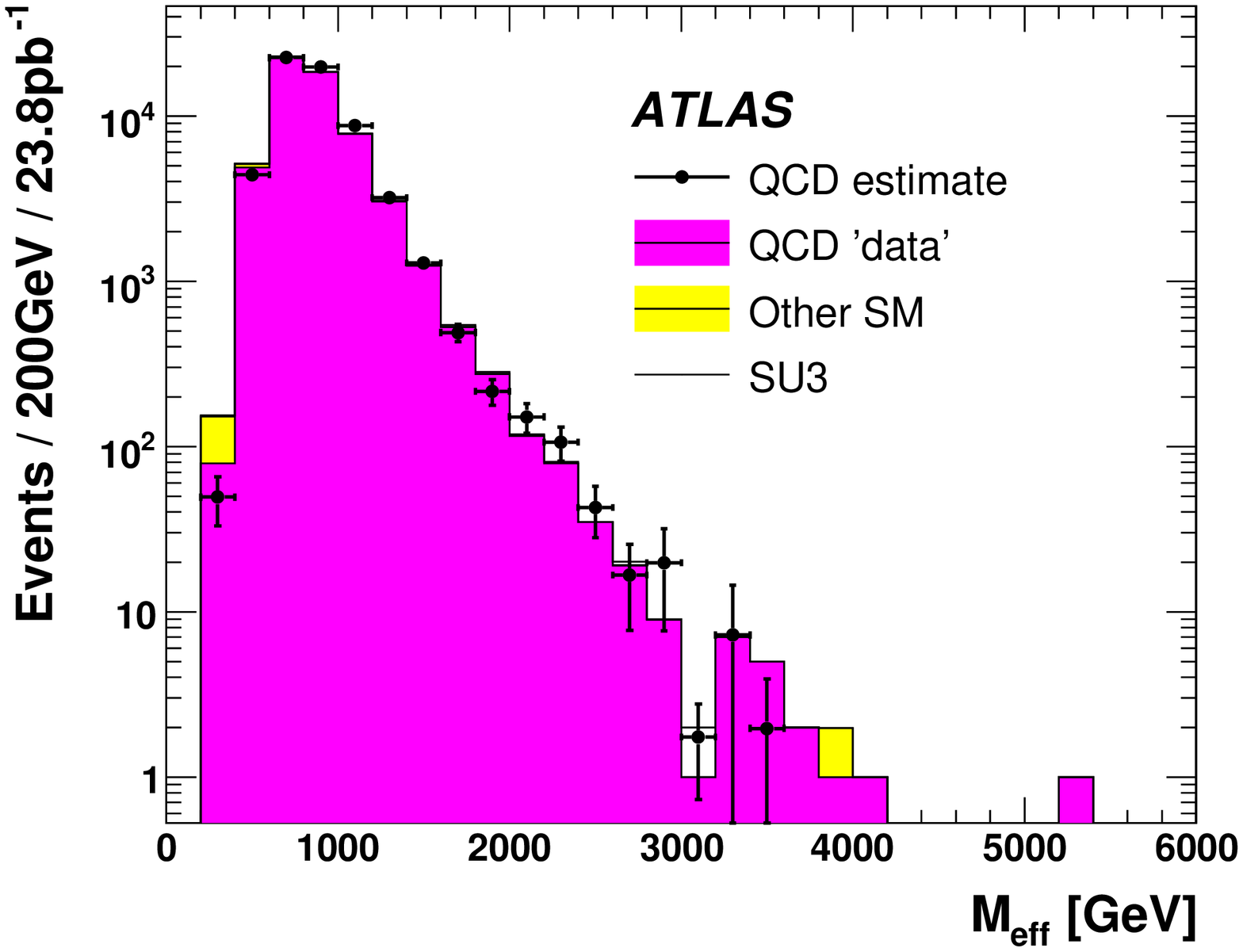}
\end{tabular}
  \caption{Left: smearing function for a jet of 250 GeV (thick line), with Gaussian and non-Gaussian
components (right and left facing hatches respectively) shown separately. Right: $\mathrm{M_{eff}}$ distribution for smeared events and GEANT4  ``data'' passing 0-lepton SUSY jet cuts \cite{CSC}. Also included for comparison are 23.8 pb$^{-1}$ of SUSY (SU3) events and the summed contribution from Z $\rightarrow$ $\nu$$\bar{\nu}$ + jets, $W$$\rightarrow$ $\ell$ $\nu$+ jets and $t$$\bar{t}$ + jets.}

\end{figure}

\section{Summary}

We examined techniques for estimating the QCD background uncertainties in searches for Supersymmetry at ATLAS.
In particular, we examined uncertainties arising from jet-energy-scale and Monte Carlo generators. We showed for example that if a 1\% jet-energy-scale uncertainty is achieved, this would affect the  $\mathrm{M_{eff}}$ uncertainty in a $\sim$ 5-10\% level. We also explored the ATLAS data-driven techniques for early-data with
minimum reliance on Monte Carlo simulations. Such techniques have the advantage to be less prone to Monte Carlo input systematics, and can, in some cases, benefit from the large statistics of the control channels used to measure the detector performance.


\begin{theacknowledgments}
This work was supported by the Federal Ministry of Education and Research (Germany).
\end{theacknowledgments}

\end{document}